\documentclass[
	aps,
	prd, 
	onecolumn, 
	a4paper, 
	10pt, 
	amsmath,
	amssymb, 
	preprintnumbers,  
	tightenlines, 
	notitlepage,
	floatfix,
	nofootinbib,
	longbibliography
	]{revtex4-1}\pdfoutput=1\pdfsuppresswarningpagegroup=1
\usepackage[utf8]{inputenc}
\usepackage[english]{babel}
\usepackage{bbm}
\usepackage{graphicx}
\usepackage{hyperref}
\usepackage{wasysym}
\usepackage{amsthm}
\usepackage{dcolumn}
	\newcolumntype{.}{D{.}{.}{13}}
	\newcolumntype{d}[1]{D{.}{.}{#1}}
\usepackage{color}	
\usepackage[all,cmtip]{xy}
\usepackage{tensor}

\graphicspath{{pics/}}

\allowdisplaybreaks[0]

\newcommand{\abs}[1]{\lvert#1\rvert}			%Absolute value
\newcommand{\hh}[1]{\left(#1\right) }			%Scaling parentheses
\newcommand{\bh}[1]{\bigl(#1\bigr) }			%big parentheses
\newcommand{\Bh}[1]{\Bigr(#1\Bigr) }			%Big parentheses
\newcommand{\bbh}[1]{\biggl(#1\biggr) }			%bigg parentheses
\newcommand{\id}[1]{\operatorname{d}\!#1}												%Differential
\renewcommand{\d}[2]{\frac{\operatorname{d}\!#1}{\operatorname{d}\!#2}}					%Derivative
\newcommand{\pd}[2]{\frac{\operatorname{\partial}\!#1}{\operatorname{\partial}\!#2}}	%Partial derivative
\newcommand{\CD}[1]{\nabla_{#1}}

\newcommand{\ii}{i}								%Imaginary unit
\newcommand{\nE}{\mathcal{E}}								%Specific energy
\newcommand{\nL}{\mathcal{L}}							%Specific angular momentum
\newcommand{\nQ}{Q}								%Specific Carter constant

\newcommand{\rmin}{r_\mathrm{min}}
\newcommand{\rmax}{r_\mathrm{max}}
\newcommand{\zmax}{z_\mathrm{max}}

\DeclareMathOperator{\sign}{sign}				%Sign

\makeatletter

\newif\if@restonecol
\newif\if@titlepage
\newif\ifiopams
\@titlepagefalse

\DeclareMathAlphabet{\bi}{OML}{cmm}{b}{it}
\DeclareMathAlphabet{\bcal}{OMS}{cmsy}{b}{n}
\renewcommand\normalsize{%
   \@setfontsize\normalsize\@xiipt{16}%
   \abovedisplayskip 12\p@ \@plus3\p@ \@minus7\p@
   \abovedisplayshortskip \z@ \@plus3\p@
   \belowdisplayshortskip 6.5\p@ \@plus3.5\p@ \@minus3\p@
   \belowdisplayskip \abovedisplayskip
   \let\@listi\@listI}
\normalsize
\renewcommand\small{%
   \@setfontsize\small\@xipt{14}%
   \abovedisplayskip 11\p@ \@plus3\p@ \@minus6\p@
   \abovedisplayshortskip \z@ \@plus3\p@
   \belowdisplayshortskip 6.5\p@ \@plus3.5\p@ \@minus3\p@
   \def\@listi{\leftmargin\leftmargini
               \topsep 9\p@ \@plus3\p@ \@minus5\p@
               \parsep 4.5\p@ \@plus2\p@ \@minus\p@
               \itemsep \parsep}%
   \belowdisplayskip \abovedisplayskip
}
\renewcommand\footnotesize{%
   \@setfontsize\footnotesize\@xpt{13}%
   \abovedisplayskip 10\p@ \@plus2\p@ \@minus5\p@
   \abovedisplayshortskip \z@ \@plus3\p@
   \belowdisplayshortskip 6\p@ \@plus3\p@ \@minus3\p@
   \def\@listi{\leftmargin\leftmargini
               \topsep 6\p@ \@plus2\p@ \@minus2\p@
               \parsep 3\p@ \@plus2\p@ \@minus\p@
               \itemsep \parsep}%
   \belowdisplayskip \abovedisplayskip
}
\renewcommand\scriptsize{\@setfontsize\scriptsize\@viiipt{9.5}}
\renewcommand\tiny{\@setfontsize\tiny\@vipt\@viipt}
\renewcommand\large{\@setfontsize\large\@xivpt{18}}
\renewcommand\Large{\@setfontsize\Large\@xviipt{22}}
\renewcommand\LARGE{\@setfontsize\LARGE\@xxpt{25}}
\renewcommand\huge{\@setfontsize\huge\@xxvpt{30}}

\if@twocolumn
\else
\fi
\def\@listI{\leftmargin\leftmargini
     \parsep=\z@
     \topsep=6\p@  \@plus3\p@ \@minus3\p@
     \itemsep=3\p@ \@plus2\p@ \@minus1\p@}
\let\@listi\@listI
\@listi
\def\@listii {\leftmargin\leftmarginii
     \labelwidth\leftmarginii
     \advance\labelwidth-\labelsep
     \topsep=3\p@ \@plus2\p@ \@minus\p@
     \parsep=\z@
     \itemsep=\parsep}
\def\@listiii{\leftmargin\leftmarginiii
     \labelwidth\leftmarginiii
     \advance\labelwidth-\labelsep
     \topsep=\z@
     \parsep=\z@
     \partopsep=\z@
     \itemsep=\z@}
\def\@listiv {\leftmargin\leftmarginiv
     \labelwidth\leftmarginiv
     \advance\labelwidth-\labelsep}
\def\@listv{\leftmargin\leftmarginv
     \labelwidth\leftmarginv
     \advance\labelwidth-\labelsep}
\def\@listvi {\leftmargin\leftmarginvi
     \labelwidth\leftmarginvi
     \advance\labelwidth-\labelsep}
\def\ps@headings{\let\@oddfoot\@empty
      \let\@evenfoot\@empty
      \def\@evenhead{{\thepage\hfil\itshape\sffamily\rightmark}}%
      \def\@oddhead{{\itshape\sffamily\leftmark}\hfil\thepage}%
      \let\@mkboth\markboth
      \let\sectionmark\@gobble
      \let\subsectionmark\@gobble}
\def\ps@myheadings{\let\@oddfoot\@empty\let\@evenfoot\@empty
    \let\@oddhead\@empty\let\@evenhead\@empty
    \let\@mkboth\@gobbletwo
    \let\sectionmark\@gobble
    \let\subsectionmark\@gobble}
\renewcommand\maketitle{\newpage}
\renewcommand{\title}{\@ifnextchar[{\@stitle}{\@ftitle}}
\def\@stitle[#1]#2{\markboth{#1}{#1}%
    \thispagestyle{myheadings}%
    \vspace*{3pc}{\exhyphenpenalty=10000\hyphenpenalty=10000 
    \Large\raggedright\noindent
    \bf\sffamily#2\par}}
\def\@ftitle#1{\markboth{#1}{#1}%
    \thispagestyle{myheadings}%
    \vspace*{3pc}{\exhyphenpenalty=10000\hyphenpenalty=10000 
    \Large\raggedright\noindent
    \bf\sffamily#1\par}}
\newcommand{\article}{\@ifnextchar[{\@sarticle}{\@farticle}}
\def\@sarticle[#1]#2#3{\markboth{#1}{#1}%
    \thispagestyle{myheadings}%
     \vspace*{.5pc}%
    {\parindent=\mathindent \bf #2\par}%
     \vspace*{1.5pc}%
    {\exhyphenpenalty=10000\hyphenpenalty=10000
     \Large\raggedright\noindent
     \bf#3\par}}%
\def\@farticle#1#2{\markboth{#2}{#2}%
    \thispagestyle{myheadings}%
     \vspace*{.5pc}%
    {\parindent=\mathindent \bf #1\par}%
     \vspace*{1.5pc}%
    {\exhyphenpenalty=10000\hyphenpenalty=10000
     \Large\raggedright\noindent
     \bf#2\par}}%
\def\review{\@ifnextchar[{\@sreview}{\@freview}}
\def\@sreview[#1]#2{\@sarticle[#1]{REVIEW ARTICLE}{#2}}
\def\@freview#1{\@farticle{REVIEW ARTICLE}{#1}}
\def\topical{\@ifnextchar[{\@stopical}{\@ftopical}}
\def\@stopical[#1]#2{\@sarticle[#1]{TOPICAL REVIEW}{#2}}
\def\@ftopical#1{\@farticle{TOPICAL REVIEW}{#1}}
\def\comment{\@ifnextchar[{\@scomment}{\@fcomment}}
\def\@scomment[#1]#2{\@sarticle[#1]{COMMENT}{#2}}
\def\@fcomment#1{\@farticle{COMMENT}{#1}}
\def\rapid{\@ifnextchar[{\@srapid}{\@frapid}}
\def\@srapid[#1]#2{\@sarticle[#1]{RAPID COMMUNICATION}{#2}}
\def\@frapid#1{\@farticle{RAPID COMMUNICATION}{#1}}
\def\note{\@ifnextchar[{\@snote}{\@fnote}}
\def\@snote[#1]#2{\@sarticle[#1]{NOTE}{#2}}
\def\@fnote#1{\@farticle{NOTE}{#1}}
\def\prelim{\@ifnextchar[{\@sprelim}{\@fprelim}}
\def\@sprelim[#1]#2{\@sarticle[#1]{PRELIMINARY COMMUNICATION}{#2}}
\def\@fprelim#1{\@farticle{PRELIMINARY COMMUNICATION}{#1}}
\renewcommand{\author}{\@ifnextchar[{\@sauthor}{\@fauthor}}
\def\@sauthor[#1]#2{\markright{#1}    % for production only
   \vspace*{1.5pc}%
   \begin{indented}%
   \item[]\normalsize\bf\sffamily\raggedright#2
   \end{indented}%
   \smallskip}
\def\@fauthor#1{%\markright{#1}         for production only
   \vspace*{1.5pc}%
   \begin{indented}%
   \item[]\normalsize\bf\sffamily\raggedright#1
   \end{indented}%
   \smallskip}
\renewcommand{\address}[1]{\begin{indented}
   \item[]\rm\raggedright #1
   \end{indented}}
\def\nosections{\vspace{30\p@ plus12\p@ minus12\p@}
    \noindent\ignorespaces}
\long\def\@makefntext#1{\parindent 1em\noindent 
 \makebox[1em][l]{\footnotesize\rm$\m@th{\fnsymbol{footnote}}$}%
 \footnotesize\rm #1}
\def\@makefnmark{\hbox{${\fnsymbol{footnote}}\m@th$}}
\def\@thefnmark{\fnsymbol{footnote}}
\def\footnote{\@ifnextchar[{\@xfootnote}{\stepcounter{\@mpfn}%
       \begingroup\let\protect\noexpand
       \xdef\@thefnmark{\thempfn}\endgroup
     \@footnotemark\@footnotetext}}
\def\@xfootnote[#1]{\setcounter{footnote}{#1}%
   \addtocounter{footnote}{-1}\footnote}
\def\@fnsymbol#1{\ifcase#1\or \dagger\or \ddagger\or \S\or
   \|\or \P\or ^{+}\or ^{\tsty *}\or \sharp
   \or \dagger\dagger \else\@ctrerr\fi\relax}
\newcounter{jnl}

\def\journal{\ifnum\thejnl=0 Institute of Physics Publishing\fi
        \ifnum\thejnl=1 J. Phys.\ A: Math.\ Gen.\ \fi
        \ifnum\thejnl=2 J. Phys.\ B: At.\ Mol.\ Opt.\ Phys.\ \fi
        \ifnum\thejnl=3 J. Phys.:\ Condens. Matter\ \fi
        \ifnum\thejnl=4 J. Phys.\ G: Nucl.\ Part.\ Phys.\ \fi
        \ifnum\thejnl=5 Inverse Problems\ \fi
        \ifnum\thejnl=6 Class. Quantum Grav.\ \fi
        \ifnum\thejnl=7 Network: Comput.\ Neural Syst.\ \fi
        \ifnum\thejnl=8 Nonlinearity\ \fi
        \ifnum\thejnl=9 J. Opt. B: Quantum Semiclass. Opt.\ \fi
        \ifnum\thejnl=10 Waves Random Media\ \fi
        \ifnum\thejnl=11 J. Opt. A: Pure Appl. Opt.\ \fi
        \ifnum\thejnl=12 Phys. Med. Biol.\ \fi
        \ifnum\thejnl=13 Modelling Simul.\ Mater.\ Sci.\ Eng.\ \fi
        \ifnum\thejnl=14 Plasma Phys. Control. Fusion\ \fi
        \ifnum\thejnl=15 Physiol. Meas.\ \fi
        \ifnum\thejnl=16 Combust. Theory Modelling\ \fi
        \ifnum\thejnl=17 High Perform.\ Polym.\ \fi
        \ifnum\thejnl=18 Public Understand. Sci.\ \fi
        \ifnum\thejnl=19 Rep.\ Prog.\ Phys.\ \fi
        \ifnum\thejnl=20 J.\ Phys.\ D: Appl.\ Phys.\ \fi
        \ifnum\thejnl=21 Supercond.\ Sci.\ Technol.\ \fi
        \ifnum\thejnl=22 Semicond.\ Sci.\ Technol.\ \fi
        \ifnum\thejnl=23 Nanotechnology\ \fi
        \ifnum\thejnl=24 Measur.\ Sci.\ Technol.\ \fi
        \ifnum\thejnl=25 Plasma.\ Sources\ Sci.\ Technol.\ \fi
        \ifnum\thejnl=26 Smart\ Mater.\ Struct.\ \fi
        \ifnum\thejnl=27 J.\ Micromech.\ Microeng.\ \fi
        \ifnum\thejnl=28 Distrib.\ Syst.\ Engng\ \fi
        \ifnum\thejnl=29 Bioimaging\ \fi
        \ifnum\thejnl=30 J.\ Radiol. Prot.\ \fi
        \ifnum\thejnl=31 Europ. J. Phys.\ \fi
        \ifnum\thejnl=32 J. Opt. A: Pure Appl. Opt.\ \fi
        \ifnum\thejnl=33 New. J. Phys.\ \fi}
\def\ead#1{\vspace*{5pt}\address{E-mail: \mailto{#1}}}
\def\mailto#1{{\tt #1}}

\newif\ifletter 
\renewcommand\thesection       {\arabic{section}}

\def\@chapapp{Section}

\renewcommand\section{\@startsection {section}{1}{\z@}%
                   {-3.5ex \@plus -1ex \@minus -.2ex}%
                   {2.3ex \@plus.2ex}%
                   {\reset@font\normalsize\bfseries\sffamily\raggedright}}
\renewcommand\subsection{\@startsection{subsection}{2}{\z@}%
                   {-3.25ex\@plus -1ex \@minus -.2ex}%
                   {1.5ex \@plus .2ex}%
                   {\reset@font\normalsize\itshape\sffamily\raggedright}}
\renewcommand\subsubsection{\@startsection{subsubsection}{3}{\z@}%
                                     {-3.25ex\@plus -1ex \@minus -.2ex}%
                                     {-1em \@plus .2em}%
                                     {\reset@font\normalsize\itshape\sffamily}}
\renewcommand\paragraph{\@startsection{paragraph}{4}{\z@}%
                                    {3.25ex \@plus1ex \@minus.2ex}%
                                    {-1em}%
                                    {\reset@font\normalsize\itshape}}
\renewcommand\subparagraph{\@startsection{subparagraph}{5}{\parindent}%
                                       {3.25ex \@plus1ex \@minus .2ex}%
                                       {-1em}%
                                      {\reset@font\normalsize\itshape}}
\def\@sect#1#2#3#4#5#6[#7]#8{\ifnum #2>\c@secnumdepth
     \let\@svsec\@empty\else
     \refstepcounter{#1}\edef\@svsec{\csname the#1\endcsname. }\fi
     \@tempskipa #5\relax
      \ifdim \@tempskipa>\z@
        \begingroup #6\relax
          \noindent{\hskip #3\relax\@svsec}{\interlinepenalty \@M #8\par}%
        \endgroup
       \csname #1mark\endcsname{#7}\addcontentsline
         {toc}{#1}{\ifnum #2>\c@secnumdepth \else
                      \protect\numberline{\csname the#1\endcsname}\fi
                    #7}\else
        \def\@svsechd{#6\hskip #3\relax  %% \relax added 2 May 90
                   \@svsec #8\csname #1mark\endcsname
                      {#7}\addcontentsline
                           {toc}{#1}{\ifnum #2>\c@secnumdepth \else
                             \protect\numberline{\csname the#1\endcsname}\fi
                       #7}}\fi
     \@xsect{#5}}
\def\@ssect#1#2#3#4#5{\@tempskipa #3\relax
   \ifdim \@tempskipa>\z@
     \begingroup #4\noindent{\hskip #1}{\interlinepenalty \@M #5\par}\endgroup
   \else \def\@svsechd{#4\hskip #1\relax #5}\fi
    \@xsect{#3}}
\@beginparpenalty -\@lowpenalty
\@endparpenalty   -\@lowpenalty
\@itempenalty     -\@lowpenalty
\renewcommand\theenumi{\roman{enumi}}
\renewcommand\theenumii{\alph{enumii}}
\renewcommand\theenumiii{\arabic{enumiii}}

\renewcommand\p@enumii{(\theenumi)}
\renewcommand\p@enumiii{(\theenumi.\theenumii)}
\renewcommand\p@enumiv{(\theenumi.\theenumii.\theenumiii)}
\renewcommand\labelitemi{$\m@th\bullet$}

\renewcommand\labelitemiii{$\m@th\ast$}
\renewcommand\labelitemiv{$\m@th\cdot$}
 \renewenvironment{abstract}{%
       \vspace{16pt plus3pt minus3pt}
       \begin{indented}
       \item[]{\bfseries\sffamily \abstractname.}\quad\rm\ignorespaces} 
       {\end{indented}\if@titlepage\newpage\else\vspace{18\p@ plus18\p@}\fi}

    {\newpage\setcounter{page}{1}}
\def\appendix{\@ifnextchar*{\@appendixstar}{\@appendix}}
\def\@appendix{\eqnobysec\@appendixstar}
\def\@appendixstar{\@@par
 \ifnumbysec                         %  Added 30/4/94 to get Table A1,
 \@addtoreset{table}{section}        %  Table B1 etc if numbering by
 \@addtoreset{figure}{section}\fi    %  section
 \setcounter{section}{0}
 \setcounter{subsection}{0}
 \setcounter{subsubsection}{0}
 \setcounter{equation}{0}
 \setcounter{figure}{0}
 \setcounter{table}{0}
 \def\thesection{Appendix \Alph{section}}   
 \def\theequation{\ifnumbysec
      \Alph{section}.\arabic{equation}\else
      \Alph{section}\arabic{equation}\fi}  % Comment A\arabic{equation} maybe
 \def\thetable{\ifnumbysec                 % better? 15/4/95
      \Alph{section}\arabic{table}\else
      A\arabic{table}\fi}
 \def\thefigure{\ifnumbysec
      \Alph{section}\arabic{figure}\else
      A\arabic{figure}\fi}}
\def\noappendix{\setcounter{figure}{0}
     \setcounter{table}{0}
     \def\thetable{\arabic{table}}
     \def\thefigure{\arabic{figure}}}
\skip\@mpfootins = \skip\footins
\renewcommand\theequation{\arabic{equation}}
\renewcommand\thefigure{\@arabic\c@figure}
\def\fps@figure{tbp}
\def\ftype@figure{1}
\def\ext@figure{lof}
\def\fnum@figure{\figurename~\thefigure}
\renewenvironment{figure}{\footnotesize\rm\@float{figure}}%
    {\end@float\normalsize\rm}
\renewenvironment{figure*}{\footnotesize\rm\@dblfloat{figure}}{\end@dblfloat}
\renewcommand\thetable{\@arabic\c@table}
\def\fps@table{tbp}
\def\ftype@table{2}
\def\ext@table{lot}
\def\fnum@table{\tablename~\thetable}
\renewenvironment{table}{\footnotesize\rm\@float{table}}%
   {\end@float\normalsize\rm}
\renewenvironment{table*}{\footnotesize\rm\@dblfloat{table}}%
   {\end@dblfloat\normalsize\rm}
\long\def\@caption#1[#2]#3{\par\begingroup
    \@parboxrestore
    \normalsize
    \@makecaption{\csname fnum@#1\endcsname}{\ignorespaces #3}\par
  \endgroup}
\long\def\@makecaption#1#2{\vskip \abovecaptionskip 
 \begin{indented}
 \item[]{\bf #1.} #2
 \end{indented}\vskip\belowcaptionskip}
\let\@portraitcaption=\@makecaption

\DeclareOldFontCommand{\rm}{\normalfont\rmfamily}{\mathrm}
\DeclareOldFontCommand{\sf}{\normalfont\sffamily}{\mathsf}
\DeclareOldFontCommand{\tt}{\normalfont\ttfamily}{\mathtt}
\DeclareOldFontCommand{\bf}{\normalfont\bfseries}{\mathbf}
\DeclareOldFontCommand{\it}{\normalfont\itshape}{\mathit}
\DeclareOldFontCommand{\sl}{\normalfont\slshape}{\@nomath\sl}
\DeclareOldFontCommand{\sc}{\normalfont\scshape}{\@nomath\sc}
\newcommand{\pcal}{\@fontswitch{\relax}{\mathcal}}
\newcommand{\pmit}{\@fontswitch{\relax}{\mathnormal}}
\renewcommand\@pnumwidth{1.55em}
\renewcommand\@tocrmarg {2.55em}
\renewcommand\@dotsep{4.5}
\renewcommand\tableofcontents{%
    \section*{\contentsname
        \@mkboth{\uppercase{\contentsname}}{\uppercase{\contentsname}}}%
    \@starttoc{toc}%
    }
\renewcommand\l@part[2]{%
  \ifnum \c@tocdepth >-2\relax
    \addpenalty{\@secpenalty}%
    \addvspace{2.25em \@plus\p@}%
    \begingroup
      \setlength\@tempdima{3em}%
      \parindent \z@ \rightskip \@pnumwidth
      \parfillskip -\@pnumwidth
      {\leavevmode
       \large \bfseries #1\hfil \hbox to\@pnumwidth{\hss #2}}\par
       \nobreak
       \if@compatibility
         \global\@nobreaktrue
         \everypar{\global\@nobreakfalse\everypar{}}
      \fi
    \endgroup
  \fi}
\renewcommand\l@section[2]{%
  \ifnum \c@tocdepth >\z@
    \addpenalty{\@secpenalty}%
    \addvspace{1.0em \@plus\p@}%
    \setlength\@tempdima{1.5em}%
    \begingroup
      \parindent \z@ \rightskip \@pnumwidth
      \parfillskip -\@pnumwidth
      \leavevmode \bfseries
      \advance\leftskip\@tempdima
      \hskip -\leftskip
      #1\nobreak\hfil \nobreak\hbox to\@pnumwidth{\hss #2}\par
    \endgroup
  \fi}
\renewcommand\l@subsection   {\@dottedtocline{2}{1.5em}{2.3em}}
\renewcommand\l@subsubsection{\@dottedtocline{3}{3.8em}{3.2em}}
\renewcommand\l@paragraph    {\@dottedtocline{4}{7.0em}{4.1em}}
\renewcommand\l@subparagraph {\@dottedtocline{5}{10em}{5em}}
\renewcommand\listoffigures{%
    \section*{\listfigurename
      \@mkboth{\uppercase{\listfigurename}}%
              {\uppercase{\listfigurename}}}%
    \@starttoc{lof}%
    }
\renewcommand\l@figure{\@dottedtocline{1}{1.5em}{2.3em}}
\renewcommand\listoftables{%
    \section*{\listtablename
      \@mkboth{\uppercase{\listtablename}}{\uppercase{\listtablename}}}%
    \@starttoc{lot}%
    }
\let\l@table\l@figure

\renewcommand\@idxitem  {\par\hangindent 40\p@}
\renewcommand\subitem   {\par\hangindent 40\p@ \hspace*{20\p@}}
\renewcommand\subsubitem{\par\hangindent 40\p@ \hspace*{30\p@}}
\renewcommand\indexspace{\par \vskip 10\p@ \@plus5\p@ \@minus3\p@\relax}
\newcommand\contentsname{Contents}
\newcommand\listfigurename{List of Figures}
\newcommand\listtablename{List of Tables}

\renewcommand\indexname{Index}
\renewcommand\figurename{Figure}
\renewcommand\tablename{Table}

\renewcommand\abstractname{Abstract}
\renewcommand\today{\number\day\space\ifcase\month\or
  January\or February\or March\or April\or May\or June\or
  July\or August\or September\or October\or November\or December\fi
  \space\number\year}
\newcommand{\Tables}{\clearpage\section*{Tables and table captions}
    \def\fps@table{hp}\noappendix}
\newcommand{\Figures}{\clearpage\section*{Figure captions}
    \def\fps@figure{hp}\noappendix}
\newcommand{\Table}[1]{\begin{table}
  \caption{#1}
  \begin{indented}
  \lineup
  \item[]\begin{tabular}{@{}l*{15}{l}}}

\def\endTable{\end{tabular}\end{indented}\end{table}}

\newcommand{\fulltable}[1]{\begin{table}
  \caption{#1}
  \lineup
  \begin{tabular*}{\textwidth}{@{}l*{15}{@{\extracolsep{0pt plus 12pt}}l}}}
\def\endfulltable{\end{tabular*}\end{table}}

\def\thebibliography#1{\list
 {\hfil[\arabic{enumi}]}{\topsep=0\p@\parsep=0\p@
 \partopsep=0\p@\itemsep=0\p@
 \labelsep=5\p@\itemindent=-10\p@
 \settowidth\labelwidth{\footnotesize[#1]}%
 \leftmargin\labelwidth
 \advance\leftmargin\labelsep
 \advance\leftmargin -\itemindent
 \usecounter{enumi}}\footnotesize
 \def\newblock{\ }
 \sloppy\clubpenalty4000\widowpenalty4000
 \sfcode`\.=1000\relax}

\def\numrefs#1{\begin{thebibliography}{#1}}
\def\endnumrefs{\end{thebibliography}}

\def\thereferences{\list{}{\topsep=0\p@\parsep=0\p@
 \partopsep=0\p@\itemsep=0\p@\labelsep=0\p@\itemindent=-18\p@
\labelwidth=0\p@\leftmargin=18\p@
}\footnotesize\rm
\def\newblock{\ }
\sloppy\clubpenalty4000\widowpenalty4000
\sfcode`\.=1000\relax
}

\newlength{\indentedwidth}
\newdimen\mathindent
\indentedwidth=\mathindent
\newenvironment{harvard}{\list{}{\topsep=0\p@\parsep=0\p@
\partopsep=0\p@\itemsep=0\p@\labelsep=0\p@\itemindent=-18\p@
\labelwidth=0\p@\leftmargin=18\p@
}\footnotesize\rm
\def\newblock{\ }
\sloppy\clubpenalty4000\widowpenalty4000
\sfcode`\.=1000\relax}{\endlist}
\def\refs{\begin{harvard}}
\def\endrefs{\end{harvard}}
\newenvironment{indented}{\begin{indented}}{\end{indented}}
\newenvironment{varindent}[1]{\begin{varindent}{#1}}{\end{varindent}}
\def\indented{\list{}{\itemsep=0\p@\labelsep=0\p@\itemindent=0\p@
   \labelwidth=0\p@\leftmargin=\mathindent\topsep=0\p@\partopsep=0\p@
   \parsep=0\p@\listparindent=15\p@}\footnotesize\rm}

\def\varindent#1{\setlength{\varind}{#1}%
   \list{}{\itemsep=0\p@\labelsep=0\p@\itemindent=0\p@
   \labelwidth=0\p@\leftmargin=\varind\topsep=0\p@\partopsep=0\p@
   \parsep=0\p@\listparindent=15\p@}\footnotesize\rm}

\newif\ifnumbysec
\def\theequation{\ifnumbysec
      \arabic{section}.\arabic{equation}\else
      \arabic{equation}\fi}
\def\eqnobysec{\numbysectrue\@addtoreset{equation}{section}}

\newcounter{eqnval}

\def\cases#1{%
     \left\{\,\vcenter{\def\\{\cr}\normalbaselines\openup1\jot\m@th%
     \ialign{\strut$\displaystyle{##}\hfil$&\tqs
     \rm##\hfil\crcr#1\crcr}}\right.}%

\renewcommand{\qquad}{\hspace*{25pt}}
\newcommand{\tqs}{\hspace*{25pt}}

\newcommand{\tsty}{\textstyle}

\def\p@subsection     {}%
\def\p@subsubsection  {}%
\def\p@paragraph  {}%
\def\p@subparagraph  {}%
\def\;{\protect\psemicolon}
\def\psemicolon{\relax\ifmmode\mskip\thickmuskip\else\kern .3333em\fi}
\def\lineup{\def\0{\hbox{\phantom{\footnotesize\rm 0}}}%
    \def\m{\hbox{$\phantom{-}$}}%
    \def\-{\llap{$-$}}}
\newcommand{\boldarrayrulewidth}{1\p@} 
\def\bhline{\noalign{\ifnum0=`}\fi\hrule \@height  
\boldarrayrulewidth \futurelet \@tempa\@xhline}

\def\@xhline{\ifx\@tempa\hline\vskip \doublerulesep\fi
      \ifnum0=`{\fi}}

\newcommand{\fcrule}[1]{\ifnum\thetabtype=1\multispan{#1}{\hrulefill
  \hspace*{\tabcolsep}}\else\multispan{#1}{\hrulefill}\fi}
\newcommand{\ms}{\noalign{\vspace{3\p@ plus2\p@ minus1\p@}}}
\newcommand{\bs}{\noalign{\vspace{6\p@ plus2\p@ minus2\p@}}}
\newcommand{\ns}{\noalign{\vspace{-3\p@ plus-1\p@ minus-1\p@}}}
\newcommand{\es}{\noalign{\vspace{6\p@ plus2\p@ minus2\p@}}\displaystyle}%
\makeatother
\renewcommand{\rmax}{r_{1}}
\renewcommand{\rmin}{r_{2}}
\renewcommand{\zmax}{z_{1}}
\newcommand{\zp}{z_{2}}

\newcommand{\sn}{\operatorname{\mathsf{sn}}}
\newcommand{\am}{\operatorname{\mathsf{am}}}

\newcommand{\elK}{\operatorname{\mathsf{K}}}
\newcommand{\elF}{\operatorname{\mathsf{F}}}
\newcommand{\elE}{\operatorname{\mathsf{E}}}
\newcommand{\elPi}{\operatorname{\mathsf{\Pi}}}

\newcommand{\eqbreak}{\\ \nonumber &\qquad}
\newcommand{\im}{\operatorname{Im}}

\newcommand{\nK}{\mathcal{K}}

\begin{document}

\title{Analytic solutions for parallel transport along generic bound geodesics in Kerr spacetime}

\author{Maarten \surname{van de Meent}}
\ead{mmeent@aei.mpg.de}
\address{Max Planck Institute for Gravitational Physics (Albert Einstein Institute), Potsdam-Golm, Germany}
%\affiliation{Mathematical Sciences, University of Southampton, United Kingdom}

\date{\today}
\begin{abstract}
We provide analytical closed form solutions for the parallel transport along a bound geodesic in Kerr spacetime. This can be considered the lowest order approximation for the motion a spinning black hole in an extreme mass-ratio inspiral. As an illustration of the usefulness of our new found expressions we scope out the locations of spin-spin resonances in quasi-circular EMRIs. All solutions are given as functions of Mino time, which facilitates the decoupling of the equations of motion. To help physical interpretation, we also provide an analytical expression for the proper time along a geodesic as a function of Mino time.
\end{abstract} 

%\maketitle
\setlength{\parindent}{0pt} 
\setlength{\parskip}{6pt}

\section{Introduction}
The motion of a freely falling frame in general relativity is described by the parallel transport of a frame along a geodesic. This is sometimes also referred to as the motion of a test gyroscope, i.e. a test particle that carries not only about a position but also an orientation.

One situation where this becomes of interests is when a small spinning black hole orbits a much larger (possibly also spinning) black hole. Such systems --- known as extreme mass-ratio inspirals --- occur naturally in the centers of galaxies when a stellar mass black hole is captured by a supermassive black hole and form, and form a key potential source of gravitational wave for future space-based gravitational wave observatories such as LISA.

To first approximation --- ignoring all effects due to its own mass and spin --- the motion of smaller (secondary) black hole is described by the parallel transport of its spin along a (bound) geodesic in the Kerr spacetime. As long as the spin of the primary is aligned with the total angular momentum of the system, the solution is fairly simple; the test object will follow an equatorial geodesic, and the spin will precess around the plane spanned by the four-velocity and total angular momentum. 

Beyond this one can include the effects of the secondary's mass and spin order-by-order. The corrections to the trajectory are governed by the gravitational self-force (see \cite{Barack:2018yvs} for a recent review and references), and  Mathisson-Papatrou-Dixon force generated by the secondary spin (see \cite{Ruangsri:2015cvg, Warburton:2017sxk, Witzany:2019dii, Witzany:2019nml} for some recent efforts at including its effects).

In this work we are interested in the motion of the secondary spin. In recent years it has become possible to calculate the correction to the precession rate of the secondary's spin linear in the secondary's mass. First for circular orbits around a non-spinning black holes \cite{Dolan:2013roa, Bini:2014ica, Shah:2015nva, Bini:2015mza, Kavanagh:2015lva}, and later extended to eccentric orbits  \cite{Akcay:2016dku, Kavanagh:2017wot, Bini:2018aps}, and spinning-black holes \cite{Bini:2018ylh, Akcay:2017azq}.

Most of these efforts have restricted themselves to the ``easy'' case where the spin of the primary is aligned with the total angular momentum. One reason for this is the lack of easily applicable closed form solutions for the parallel transport along a generic orbit. The purpose of this work is to fill this gap.

An elegant procedure for solving the parallel transport equations along a geodesic in Kerr spacetime was set out by Marck \cite{Marck:1983} in 1983. (With in recent years some clarifications being added by Bini and collaborators \cite{Bini:2017slb, Bini:2018zfu}). This procedure effectively reduces the parallel transport equations to a single differential equation for the precession angle given a solution for the geodesic equation.

Closed form analytic solutions for bound geodesics in Kerr spacetime were derived by Fujita and Hikida \cite{Fujita:2009bp} in 2009. This paper takes their method an extends it to a solution for Marck's equation for parallel transport.

The plan of this paper is as follows. In section~\ref{sec:geodesics} we review the analytic solution for bound geodesics found by Fujita and Hikida \cite{Fujita:2009bp}. Along the way we establish most of the conventions and notations that we will need later. We also repackage the solutions of \cite{Fujita:2009bp} in much more compact and easier to use way. Section~\ref{sec:pt} then introduces Marck's formalism for solving the parallel transport equations, and gives analytical solutions following the procedures of \cite{Fujita:2009bp}.
In section~\ref{sec:resonances} we illustrate the usefulness of our new found expressions by scoping out the locations of spin-spin resonances in quasi-circular EMRIs. As a bonus result, \ref{app:propertime} gives the analytical solution of the evolution of proper time as a function of Mino time along a geodesic. 

The equations in this paper are given in geometric units such that $G = c =1$. \ref{app:elliptic} establishes our conventions for elliptic functions.

\section{Geodesic equations}\label{sec:geodesics}
The metric for a Kerr black hole of unit mass ($M=1$) and spin $a$ in (modified) Boyer-Lindquist coordinates is given by
\begin{align}
\label{eq:kerr}
\id{s}^2 &= 
-\bh{1 - \frac{2r}{\Sigma}}\id{t}^2 
+ \frac{\Sigma}{\Delta} \id{r}^2
+ \frac{\Sigma}{1-z^2} \id{z}^2
\eqbreak
+ \frac{1-z^2}{\Sigma} \bh{2a^2 r (1-z^2)+(a^2+r^2)\Sigma}\id\phi^2
- \frac{4ar(1-z^2)}{\Sigma}\id{t}\id\phi,
\end{align}
where $z$ is related to the usual polar Boyer-Lindquist coordinate $\theta$ by $z = \cos\theta$. Furthermore,
\begin{align}
\Delta 	&:= r(r-2)+a^2\\
		&= (r-r_{+})(r-r_{-}), \quad\text{and}\\
\Sigma &:= r^2 + a^2 z^2.
\end{align}
The locations of the inner and outer event horizon are denoted,
\begin{equation}
	r_\pm = 1\pm\sqrt{1-a^2}.
\end{equation}

Solving the geodesic equations is aided by the existence of four constants of motion. The first is given by the norm of the 4-velocity $u^\mu$, which is set to -1 by the normalization of the proper time $\tau$. The second and third are the (specific) energy $\nE$ and angular momentum $\nL$, given by the time translation and rotational Killing symmetries of Kerr spacetime,
\begin{align}
\nE &:= -u^\mu g_{\mu\nu}\hh{\pd{}{t}}^\nu,\quad\text{and} \\
\nL &:= u^\mu g_{\mu\nu}\hh{\pd{}{\phi}}^\nu.
\end{align}

Finally, Carter \cite{Carter:1968rr} showed that there exist a fourth constant of motion, $\nQ$, related to the existence of a Killing tensor $\mathcal{K}_{\mu\nu}$, which (given as the ``square'' of the Killing-Yano tensor $\mathcal{F}$) reads,
\begin{align}
\mathcal{K}_{\mu\nu} &:=\mathcal{F}_{\mu\alpha}\mathcal{F}^{\alpha}_{\phantom{\alpha}\nu},\quad\text{where}\\
\mathcal{F}&:= 2a z \id{r}\wedge\Bh{\id{t}-a(1-z^2)\id{\phi}} + 2ar \id{z}\wedge\Bh{\id{t}-\frac{r^2+a^2}{a}\id\phi}.
\end{align}
We define the the Carter constant as
\begin{align}
\nQ &:= u^\mu \mathcal{K}_{\mu\nu} u^\nu -(\nL-a\nE)^2.
\end{align}

Using the constants of motion the geodesic equations in Kerr spacetime can be written in first order form,
\begin{subequations}\label{eq:eoms}
\begin{align}
\hh{\d{r}{\lambda}}^2 
&= \hh{\nE(r^2+a^2)-a\nL}^2 - \Delta\hh{r^2+(a\nE-\nL)^2+\nQ}\label{eq:radeq} \\	
&= (1-\nE^2)(\rmax-r)(r-\rmin)(r-r_3)(r-r_4), \label{eq:radeqpol}
\\
\hh{\d{z}{\lambda}}^2 
&=
\nQ-z^2\Bh{a^2 (1-\nE^2)(1-z^2) + \nL^2 +\nQ} \label{eq:poleq} \\
&= (z^2-\zmax^2)\hh{a^2(1-\nE^2)z^2-\zp^2},\label{eq:poleqpol}
\\
\d{t}{\lambda} &=\frac{r^2+a^2}{\Delta}\bh{\nE(r^2+a^2)-a\nL} -a^2\nE(1-z^2)+a\nL,\text{ and}
\\ 
\d{\phi}{\lambda} &= \frac{a}{\Delta}\hh{\nE(r^2+a^2)-a\nL}+\frac{\nL}{1-z^2}-a\nE,
\end{align}
\end{subequations}
where we have introduced the Mino(-Carter) time parameter $\lambda$ \cite{Mino:2003yg}, defined by
\begin{equation}
\id\tau = \Sigma \id\lambda,
\end{equation}
in order to decouple the radial and polar equations.

In this paper we identify bound orbits in Kerr spacetime by the locations of the radial turning points $\rmax$ and $\rmin$, and the polar turning point $\zmax$. The remaining nodes in Eqs.~\eqref{eq:radeqpol} and \eqref{eq:poleqpol} can be found by comparing \eqref{eq:radeqpol} and \eqref{eq:poleqpol} to \eqref{eq:radeq} and \eqref{eq:poleq} and finding the zeroes \cite{Fujita:2009bp},
\begin{align}
r_3 &= \frac{1}{1-\nE^2}-\frac{\rmax+\rmin}{2}
+\sqrt{
	\hh{\tfrac{\rmax+\rmin}{2}-\tfrac{1}{1-\nE^2}}^2-\tfrac{a^2\nQ}{\rmax\rmin(1-\nE^2)}
}, \\
r_4 &= \frac{a^2\nQ}{\rmax\rmin r_3(1-\nE^2)},\quad\text{and}\\
\zp &= \sqrt{a^2(1-\nE^2)+\frac{\nL^2}{1-\zmax^2}}.
\end{align}
Note that our convention for the second polar root $\zp$ given here (and in Eq.~\ref{eq:poleqpol}) differs from the definitions given in (among others) \cite{Fujita:2009bp} and \cite{Schmidt:2002qk}. The advantage of this convention is that allows easy evaluation of the $a\to0$ limit.

The equations for the remaining roots can be inverted to obtain $\nE$, $\nL$, and $\nQ$ in terms of $\rmax$, $\rmin$, and $\zmax$ \cite{Schmidt:2002qk}. For convenience (and unity of notation) the explicit expression are given in \ref{app:ELQ}.
\subsection{Trajectories}
The geodesic equations \eqref{eq:eoms} can be solved as functions of Mino time in closed form using elliptic functions \cite{Fujita:2009bp,Kraniotis:2004cz,Hackmann:2008zz,Hackmann:2010tqa,Hackmann:2010zz}. We here repeat the explicit solutions for bound geodesics given by Fujita and Hikida \cite{Fujita:2009bp}, in a much simplified form.

The solutions given in \cite{Fujita:2009bp} are given in a piecewise manner, making them look more complicated then they are. By applying some of the standard identies for elliptic functions their solutions can be rewritten in a form that applies at all times. That this should be true can be seen from two simple observations: 1) the solutions should be analytic functions of Mino time, 2) elliptic functions are analytic functions of their arguments. Consequently, if we are given an analytic expression of the solution on some interval, then the analytic extension of that expression should give the solution everywhere.

The solutions of the radial equation \eqref{eq:radeq} and polar equation \eqref{eq:poleq} are given by,
\begin{align}
r(q_r) &= \frac{r_3 (\rmax-\rmin)\sn^2(\frac{\elK(k_r)}{\pi} q_r|k_r)-\rmin(\rmax-r_3)}{(\rmax-\rmin)\sn^2(\frac{\elK(k_r)}{\pi} q_r|k_r)-(\rmax-r_3)},\quad\text{and}
\\
z(q_z) &= \zmax \sn(\elK(k_z)\frac{2q_z}{\pi}|k_z),
\end{align}
where $\sn$ is Jacobi elliptic sine function, $K$ is complete elliptic integral of the first kind,
\begin{align}
k_r &:=\frac{(\rmax-\rmin)(r_3-r_4)}{(\rmax-r_3)(\rmin-r_4)},\quad\text{and}
\\
k_z &:= a^2 (1-\nE^2)\frac{\zmax^2}{\zp^2}.
\end{align}
The solutions are $2\pi$ periodic in the radial and polar phases, $q_r$ and $q_z$, which evolve linearly with Mino time
\begin{align}
q_r &:= \Upsilon_r \lambda + q_{r,0},\label{eq:qr}\\
q_z &:= \Upsilon_z \lambda + q_{z,0},\label{eq:qz}
\end{align}
with ``frequencies''
\begin{align}
\Upsilon_r  &= \frac{\pi}{2\elK(k_r)}\sqrt{(1-\nE^2)\hh{\rmax-r_3}\hh{\rmin-r_4}},\quad\text{and}\\
\Upsilon_z  &= \frac{\pi \zp}{2\elK(k_z)}.
\end{align}
Furthermore with adopted the convention that $q_r=0$ corresponds to the periapsis ($r=\rmin$), of the radial motion, and $q_z=0$ corresponds to up going node ($z=0, z'>0$) of the polar motion.\footnote{Note that the convention for the polar phase $q_z$ differs from the one used by the author in \cite{vandeMeent:2017bcc} by a shift $\pi/2$.}

The solutions for $t$ and $\phi$ are given by
\begin{align}
t(q_t,q_r,q_z) &= q_t + t_r(q_r)+t_z(q_z),\\
t_r(q_r) &:=	\tilde{t}_r\Bh{\am\bh{\elK(k_r)\frac{q_r}{\pi}\big|k_r}}, -\frac{\tilde{t}_r(\pi)}{2\pi}q_r
\\
t_z(q_z) &:=	\tilde{t}_z\Bh{\am\bh{\elK(k_z)\frac{2q_z}{\pi}\big|k_z}} -\frac{\tilde{t}_z(\pi)}{\pi}q_z
\end{align}
and
\begin{align}
\phi(q_\phi,q_r,q_z) &= q_\phi + \phi_r(q_r) + \phi_z(q_z),\\
\phi_r(q_r) &:=	\tilde{\phi}_r\Bh{\am\bh{\elK(k_r)\frac{q_r}{\pi}\big|k_r}}, -\frac{\tilde{\phi}_r(\pi)}{2\pi}q_r
\\
\phi_z(q_z) &:=	\tilde{\phi}_z\Bh{\am\bh{\elK(k_z)\frac{2q_z}{\pi}\big|k_z}}, -\frac{\tilde{\phi}_z(\pi)}{\pi},
\end{align}
where $\am$ is the amplitude for the Jacobi elliptic functions, and

\begin{align}
\tilde{t}_r(\xi_r) &:= \frac{\nE(\rmin-r_3)}{\sqrt{(1-\nE^2)(\rmax-r_3)(\rmin-r_4)}}\Bh{
	(4+\rmax+\rmin+r_3+r_4)\elPi(h_r;\xi_r|k_r)\\
\nonumber	&\quad
	-\frac{4}{r_{+}-r_{-}}\hh{
		\frac{r_{+}(4-a \nL/\nE)-2a^2}{(\rmin-r_{+})(r_3-r_{+})}\elPi(h_{+};\xi_r|k_r)- (+\leftrightarrow -)}\\
\nonumber	&\quad
	+\frac{(\rmax-r_3)(\rmin-r_4)}{\rmin-r_3}\bh{\elE(\xi_r|k_r)-h_r\frac{\sin\xi_r\cos\xi_r\sqrt{1-k_r\sin^2\xi_r}}{1-h_r\sin^2\xi_r}}
},
\\
\tilde{t}_z(\xi_z) &:= -\frac{\nE}{1-\nE^2}\zp \elE(\xi_z|k_z),\\
\tilde{\phi}_r(\xi_r) &:= -\frac{2a\nE(\rmin-r_3)}{(r_{+}-r_{-})\sqrt{(1-\nE^2)(\rmax-r_3)(\rmin-r_4)}}
\eqbreak\times
\Bh{
		\frac{2r_{+}-a \nL/\nE}{(\rmin-r_{+})(r_3-r_{+})}\elPi(h_{+};\xi_r|k_r)- (+\leftrightarrow -)
},\\
\tilde{\phi}_z(\xi_z) &:= -\frac{\nL}{\zp} \elPi(\zmax^2;\xi_z|k_z),
\end{align}
where $\elE$ and $\elPi$ are elliptic integrals of the second and third kind,
\begin{align}
h_r &:= \frac{\rmax-\rmin}{\rmax-r_3},\\
h_{\pm} &:=h_r  \frac{r_3-r_{\pm}}{\rmin-r_{\pm}},
\end{align}
and $(+\leftrightarrow -)$ denotes that the preceding term is to be repeated with the $+$ and $-$ symbols exchanged.

The ``phases'', $q_t$ and $q_\phi$, represent the secularly growing linear parts of the solutions
\begin{align}
q_t &:= \Upsilon_t \lambda + q_{t,0},\quad\text{and}\\
q_\phi &:= \Upsilon_\phi \lambda + q_{\phi,0},
\end{align}
with
\begin{align}
\Upsilon_t	&= \tilde{\Upsilon}_{t,r} + \tilde{\Upsilon}_{t,z},\\
\Upsilon_\phi	&= \tilde{\Upsilon}_{\phi,r} + \tilde{\Upsilon}_{\phi,z},
\end{align}
and
\begin{align}
\tilde{\Upsilon}_{t,r} &:= (4+a^2)\nE +\nE\bbh{
\eqbreak
	\frac{1}{2}\Bh{
		\hh{4+\rmax+\rmin+r_3}r_3 -\rmax\rmin 
		+(\rmax-r_3)(\rmin-r_4)\frac{\elE(k_r)}{\elK(k_r)}
		\eqbreak\quad
		+\hh{4+\rmax+\rmin+r_3+r_4}(\rmin-r_3)\frac{\elPi(h_r|k_r)}{\elK(k_r)}		
	}\eqbreak
	+\frac{2}{r_{+}-r_{-}}\Bh{ 
		\frac{(4-a\nL/\nE)r_{+} -2 a^2}{r_3-r_{+}} \bh{1-\frac{\rmin-r_3}{\rmin-r_{+}}\frac{\elPi(h_{+}|k_r)}{\elK(k_r)}}
		- (+\leftrightarrow -)
	}
},
\\
\tilde{\Upsilon}_{t,z} &:=-a^2\nE +\frac{\nE\nQ}{(1-\nE^2)\zmax^2}\Bh{
	1-\frac{\elE(k_z)}{\elK(k_z)}
},
\\
\tilde{\Upsilon}_{\phi,r} &:= \frac{a}{r_{+}-r_{-}}\bbh{
	\frac{2\nE r_{+}-a\nL}{r_3-r_{+}}\Bh{1-\frac{\rmin-r_3}{\rmin-r_{+}} \frac{\elPi(h_{+}|k_r)}{\elK(k_r)} }
	- (+\leftrightarrow -)
},
\\
\tilde{\Upsilon}_{\phi,z} &:= \frac{\nL}{\elK(k_z)}\elPi(\zmax^2|k_z).
\end{align}

One useful aspect of these solutions is that they give explicit closed form expressions for the Mino time frequencies $\Upsilon_i$. From these one easily obtains the frequencies with respect to coordinate (or Killing) time, $\Omega_i$ by taking the ratio with $\Upsilon_t$,
\begin{equation}
\Omega_i = \frac{\Upsilon_i}{\Upsilon_t}.
\end{equation}
Moreover, the frequencies with respect to proper time are obtained as
\begin{equation}
\omega_i = \frac{\Upsilon_i}{\Upsilon_\tau}.
\end{equation}
This requires the average linear increase of proper time with Mino time $\Upsilon_\tau$, which is given in \ref{app:propertime}.

%%%%%%%%%%%%%%%%%%%%%%%%%%%%%%%%%%%%%%%%%%%%%%%%%%%%%%%%%%%%%%%%%%%%%%%%%%%%%%%%%%%%%%%%%%%%%%%%%%%%
\section{Parallel transport}\label{sec:pt}
We want to find a tetrad $(e_{i})_\mu$ of (co-)vectors that is parallel transported along a generic bound geodesic in Kerr spacetime, 
\begin{align}\label{eq:pt}
u^\alpha \CD{\alpha} (e_{i})_\mu =  \d{}{\tau} (e_{i})_\mu - \Gamma^\alpha_{\beta\mu}u^\beta (e_{i})_\alpha =0.
\end{align}
In the following tetrad indices are denoted by Roman letters and run from 0 to 3, and spacetime indices are denoted by Greek letters.

In 1983, Marck \cite{Marck:1983} gave a general procedure for finding such a tetrad given a solution to the geodesic equation in Kerr. We here follow his approach.
 
A geodesic, by definition, parallel transports its own 4-velocity, $u_\mu$. We can therefore use it as the first leg of our tetrad,
\begin{align}
(e_{0})_\mu &:= u_\mu = (-\nE, \frac{1}{\Delta} \d{r}{\lambda},\frac{1}{1-z^2}\d{z}{\lambda},\nL).
\end{align}
Moreover, it follows directly from the defining property of a Killing-Yano tensor, $\CD{(\alpha}\mathcal{F}_{\mu)\nu}=0$, that $\mathcal{F}_{\mu\alpha}u^\alpha$ is also parallel transported along a geodesic. This quantity is sometimes interpreted as the total (specific) orbital angular momentum of the geodesic. For us, it serves to construct the last leg of our tetrad,
\begin{align}
(e_{3})_\mu &:= \frac{\mathcal{F}_{\mu\alpha}u^\alpha}{\sqrt{\nK}} 
= \frac{1}{\sqrt{\nK}}\begin{pmatrix}
-a\frac{r\d{z}{\lambda}+z\d{r}{\lambda}}{\Sigma}\\
a z\frac{(r^2+a^2)\nE-a\nL}{\Delta}\\
a r \nE-\frac{r \nL}{1-z^2}\\
\frac{a^2 z(1-z^2)\d{r}{\lambda}+r(r^2+a^2)\d{z}{\lambda}}{\Sigma}
\end{pmatrix},
\end{align}
where
\begin{align}
\nK 	&:= u^\alpha\nK_{\alpha\beta}u^\beta = \nQ+(a\nE-\nL)^2.
\end{align}
The remaining two tetrad legs must lie in the plane perpendicular to the (co-)vectors $(e_{0})_\mu$ and $(e_{3})_\mu$. Following Marck~\cite{Marck:1983} we construct an orthonormal basis for this plane,
\begin{align}
(\tilde{e}_{1})_\mu &:= \frac{1}{\sqrt{\nK}}\begin{pmatrix}
\frac{- \Xi r \d{r}{\lambda} +\frac{a^2 z}{\Xi}\d{z}{\lambda}}{\Sigma}\\
\Xi r\frac{(r^2+a^2)\nE-a\nL}{\Delta}\\
-\frac{a z}{\Xi}\bh{a\nE-\frac{\nL}{1-z^2}}\\
a\frac{\Xi^2 r(1-z^2)\d{r}{\lambda}-z(r^2+a^2)\d{z}{\lambda}}{\Xi\Sigma}
\end{pmatrix},\quad\text{and}
\\
(\tilde{e}_{2})_\mu &:=\begin{pmatrix}
\frac{\nE}{\Xi}-\frac{(1-\Xi^2)\bh{(r^2+a^2)\nE-a\nL}}{\Xi\Sigma}\\
-\frac{\Xi}{\Delta}\d{r}{\lambda}\\
-\frac{1}{\Xi(1-z^2)}\d{z}{\lambda}\\
-\Xi\nL-\frac{(1-\Xi^2)(r^2+a^2)\hh{\nL-a(1-z^2)\nE}}{\Xi\Sigma}
\end{pmatrix},
\end{align}
with
\begin{align}
	\Xi		&:= \sqrt{\frac{\nK-a^2z^2}{\nK+r^2}}.
\end{align}
It follows that the remaining two legs of our parallel propagated tetrad must be of the form, 
\begin{align}
(e_{1})_\mu &:=\phantom{-} \cos\psi(\lambda)(\tilde{e}_{1})_\mu  + \sin\psi(\lambda)(\tilde{e}_{2})_\mu,\quad\text{and} \\
(e_{2})_\mu &:= -\sin\psi(\lambda)(\tilde{e}_{1})_\mu + \cos\psi(\lambda)(\tilde{e}_{2})_\mu. 
\end{align}
The requirement that $(e_{1})_\mu$ and $(e_{2})_\mu$ satisfy the parallel transport equation \eqref{eq:pt}, reduces (after some straightforward but tedious algebra) to a single first order differential equation for $\psi(\lambda)$,
\begin{align}
	\d{\psi}{\lambda} = \sqrt{\nK}\hh{ \frac{(r^2+a^2)\nE-a\nL}{\nK+r^2} +a\frac{\nL-a (1-z^2)\nE}{\nK-a^2z^2}}.
\end{align}
Since the right hand side of this equation can be written as the sum of two rational functions, in $r$ and $z$ respectively, it can be solved in terms standard elliptic functions. Following the same general procedure as used in \cite{Fujita:2009bp} to solve the equations for $t$ and $\phi$, we find,
\begin{align}
\psi(q_\psi,q_r,q_z) &= q_\psi + \psi_r(q_r)+\psi_z(q_z)\\
\psi_r(q_r) &:=	\tilde{\psi}_r\Bh{\am\bh{\elK(k_r)\frac{q_r}{\pi}\big|k_r}} -\frac{\tilde{\psi}_r(\pi)}{2\pi}q_r
\\
\psi_z(q_z) &:=	\tilde{\psi}_z\Bh{\am\bh{\elK(k_z)\frac{2q_z}{\pi}\big|k_z}} -\frac{\tilde{\psi}_z(\pi)}{\pi}q_z
\end{align}
with
\begin{align}
	\tilde{\psi}_r(\xi_r) &:= \frac{2(\rmin-r_3)\bh{(\nK-a^2)\nE+a\nL}}{(\nK+\rmin^2)(\nK+r_3^2)\sqrt{(1-\nE^2)(\rmax-r_3)(\rmin-r_4)}}
	\eqbreak
	\times \im\bbh{
	(r_2+ \ii \sqrt{\nK})(r_3+ \ii \sqrt{\nK})\elPi(h_\psi;\xi_r|k_r)
	},
	\\
	\tilde{\psi}_z(\xi_z) &:=-\frac{(a^2-\nK)\nE-a\nL}{\sqrt{\nK}\zp}\elPi(\frac{a^2\zmax^2}{\nK};\xi_z|k_z),
	\end{align}
and
\begin{align}
h_\psi := h_r \frac{r_3 -\ii \sqrt{\nK}}{r_2 -\ii \sqrt{\nK}}.
\end{align}
The phases  $q_r$,$q_z$, and $q_\psi$ are functions of Mino time given by \eqref{eq:qr}, \eqref{eq:qz}, and
\begin{equation}
q_\psi = \Upsilon_\psi \lambda +q_{\psi,0}.
\end{equation}
Here $\Upsilon_\psi$ is the (Mino time) precession frequency of the tetrad and is given by
\begin{align}
\Upsilon_\psi &= \Upsilon_{\psi,r} + \Upsilon_{\psi,z} \\
\Upsilon_{\psi,r} &:= \frac{\sqrt{\nK}\hh{(r_3^2+a^2)\nE-a\nL}}{r_3^2+\nK}\eqbreak
+\frac{(\rmin-r_3)\hh{(\nK-a^2)\nE+a\nL}}{(\rmin^2+\nK)(r_3^2+\nK)} \im\Bh{(r_2+\ii\sqrt{\nK})(r_3+\ii\sqrt{\nK})\frac{\elPi(h_\psi|k_r)}{\elK(k_r)}}
\\
\Upsilon_{\psi,z} &:=-\nE\sqrt{\nK}+\frac{(\nK-a^2)\nE+a\nL}{\sqrt{\nK}}\frac{\elPi(\frac{a^2\zmax^2}{\nK}|k_z)}{\elK(k_z)}.
\end{align}

We thus find a complete closed form analytic solution for the parallel propagation of a tetrad along a generic bound geodesic in Kerr spacetime. A \textsc{Mathematica} implementation of these solutions is included in the \textsf{KerrGeodesics} package of the \textit{Black Hole Perturbation Toolkit} \cite{BHPToolkit}.

\section{Application: spin-spin resonances for spherical orbits}\label{sec:resonances}

One corollary of the analytical solution of parallel transport along a bound geodesic (that was already known from Marck's work \cite{Marck:1983}), is that parallel transport introduces only one new independent frequency, $\Upsilon_\psi$, to the three frequencies, $(\Upsilon_r, \Upsilon_z, \Upsilon_\phi)$, which characterize a bound geodesic. Consequently, any time variable effects in the dynamics of a test rigid body around a black hole has a frequency spectrum that consists of integer combinations of $\Omega_r$, $\Omega_z$, $\Omega_\phi$, and $\Omega_\psi$.

It can occur that for some bound orbit one such combination vanishes. In such a case an effect that is normally oscillatory in nature becomes constant, allowing it to grow secularly over many orbits. Such a situation is known as an (orbital) resonance. The effects of resonances involving the orbital frequencies $\Omega_r$, $\Omega_z$, and $\Omega_\phi$ in extreme mass-ratio inspirals (EMRIs) have been subject of extensive study~\cite{Flanagan:2010cd,Flanagan:2012kg,Ruangsri:2013hra,Brink:2015roa,Gair:2011mr,Berry:2016bit,Mihaylov:2017qwn,vandeMeent:2013sza,vandeMeent:2014raa,Hirata:2010xn,Bonga:2019ycj}. The appearance of a fourth frequency $\Omega_\psi$ allows for the occurrence of new types of resonances.

Having a closed from analytical expression for $\Upsilon_\psi$ (and the other frequencies), is helpful in finding out where and when such resonance can occur. In this section we demonstrate the utility of our expressions by mapping out the locations of resonances between the polar $\Omega_z$ frequency and the precession frequency  $\Omega_\psi$ ($z\psi$-resonances) for inclined circular (a.k.a. spherical) orbits.\footnote{This restriction is pure for the sake of easy of presentation. There is no technical obstruction to finding the generic orbit resonances.} 

%%%%%%%%%%%%%%%%%%%%%%%%%%%%%%%%%%%%%%%%%%%%%%%%%%%%%%%%%%%%%%%%%%%%%%%%%%%%%%%%%%%%%%%%
\begin{figure}[t]
	\includegraphics[width=.7\columnwidth]{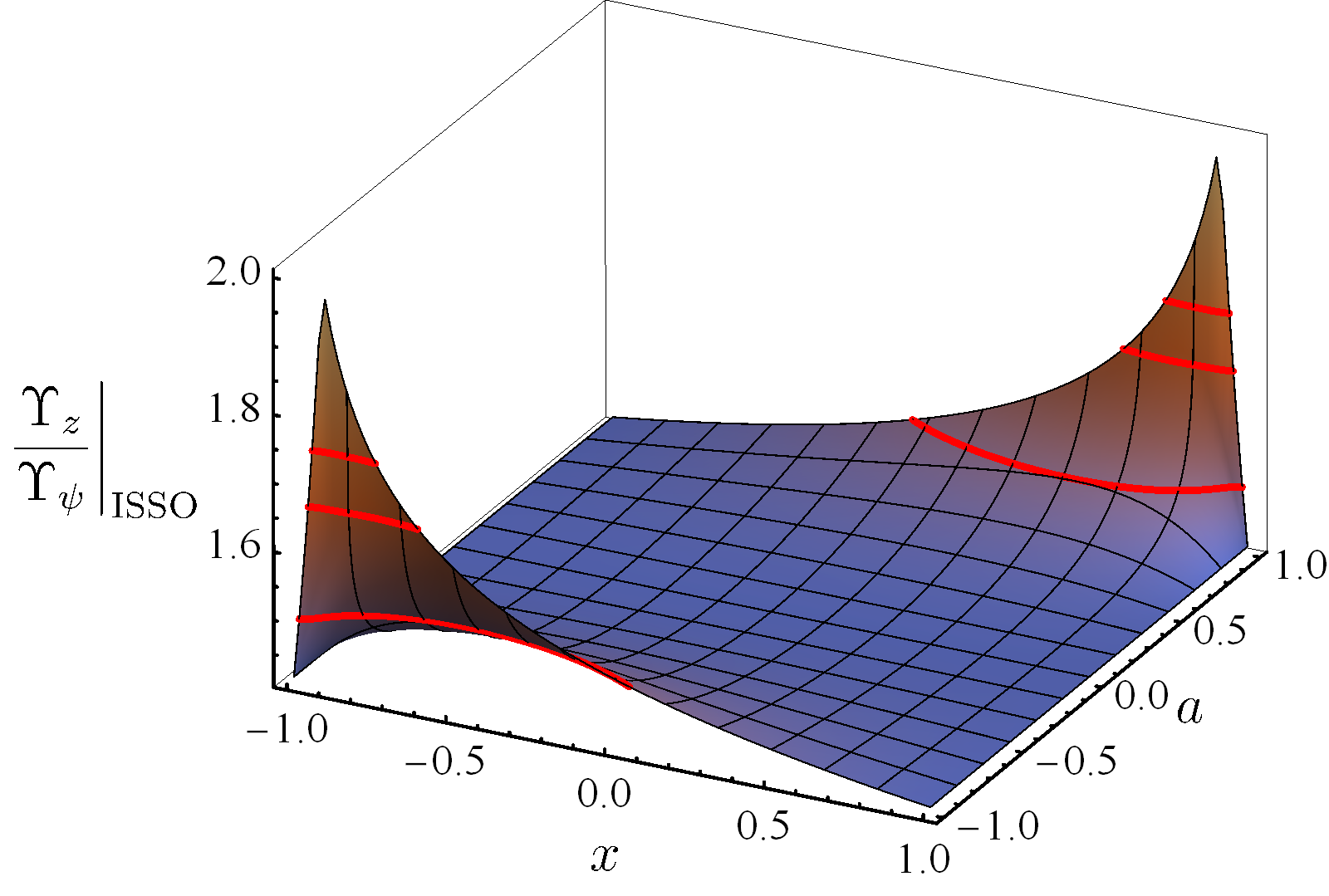}
	\caption{The ratio of  the polar frequency $\Upsilon_z$ to the precession frequency $\Upsilon_\psi$ evaluated at the last stable orbit. The red bands indicate the 3:2, 5:3, and 7:4 resonances. The maximum value of 2 is achieved at $(x,a) = (\pm\sqrt{2(\sqrt{2}-1)} ,\pm 1)$. The minimum value of $\sqrt{2}$ is attained when either $a=0$ or $\abs{x}=1$.
	}
	\label{fig:ISSOrat}
\end{figure}
%%%%%%%%%%%%%%%%%%%%%%%%%%%%%%%%%%%%%%%%%%%%%%%%%%%%%%%%%%%%%%%%%%%%%%%%%%%%%%%%%%%%%%%%
\begin{figure}[t]
	\includegraphics[width=.7\columnwidth]{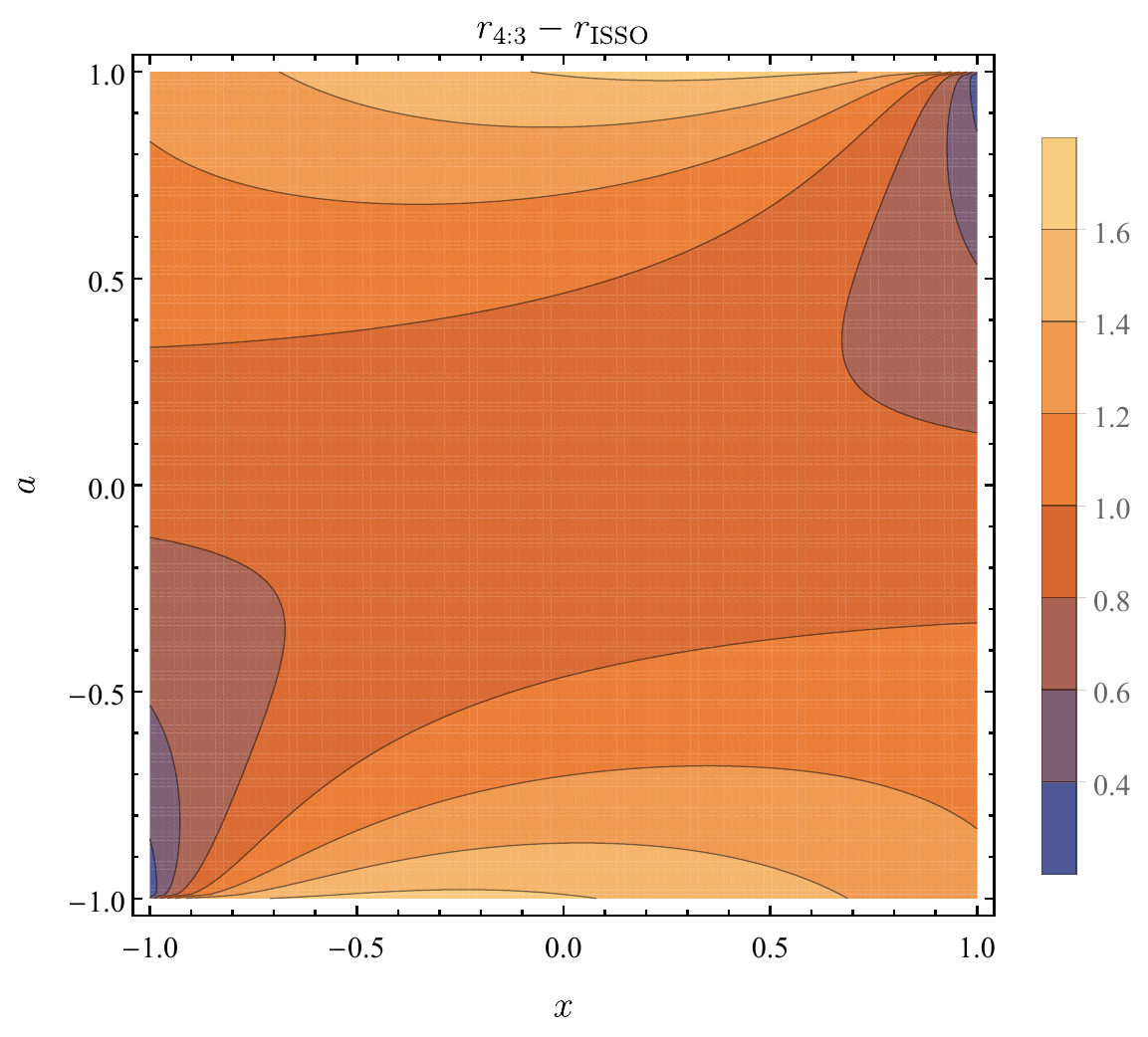}
	\caption{The location of the 4:3 $z\psi$-resonance relative to the location of the innermost stable spherical orbit. This distance is typical of the order of one gravitational radius, meaning the  4:3 $z\psi$-resonance is achieved well before the plunge phase.
	}
	\label{fig:zpsi43}
\end{figure}
%%%%%%%%%%%%%%%%%%%%%%%%%%%%%%%%%%%%%%%%%%%%%%%%%%%%%%%%%%%%%%%%%%%%%%%%%%%%%%%%%%%%%%%%

Note that the coordinate time frequencies $\Omega_i$ satisfy a resonance condition if and only if the corresponding Mino time frequencies $\Upsilon_i$ do so as well. Hence, to look for resonances, we can simply look for (low order) rational ratios $\Upsilon_z/\Upsilon_\psi$.

For resonances involving the $\Omega_r$ it is clear that on a trajectory connecting the weak field limit to the last stable orbit $\Upsilon_r/\Upsilon_i$ must pass through all possible resonant ratios between 0 and 1, because by definition $\Upsilon_r=0$ at the last stable orbit and $\Upsilon_r=\Upsilon_i$ in the weak field. However, for  $\Upsilon_z/\Upsilon_\psi$ there is no such guarantee. It is still true that $\Upsilon_z=\Upsilon_\psi$ in the weak field limit. Moreover, we observe empirically that the ratio $\Upsilon_z/\Upsilon_\psi$ increases monotonically with $1/r$ at fixed $a$ and inclination $\zmax$.

Hence, we can determined the range of resonance that can occur for spherical orbits by first determining the ratio $\Upsilon_z/\Upsilon_\psi$ at the innermost stable spherical orbit (ISSO). Figure~\ref{fig:ISSOrat} displays this ratio as a function of the background spin $a$ and the inclination parameter $x := \pm \sqrt{1-\zmax^2}$ (with the sign depend on the ``sense'' of the orbit, i.e. prograde equatorial orbits have $x=1$ and retrograde equatorial orbits $x=-1$.) The minimal value of this ratio attained when $a=0$ or $\abs{x}=1$, in which case our analytic expressions tell us that the ratio is exactly $\sqrt{2}$. From Fig.~\ref{fig:ISSOrat} we also see that the maximum ratio is attained for $\abs{a}=1$. To find the exact value of this ratio, we note that the location of the ISSO for an extremal Kerr black hole is determined by the relations
\begin{equation}
\label{eq:extISSO}
\zmax^2 =\frac{r^2\hh{3+3\sqrt{r}-2\sqrt{3+2\sqrt{r}+3r}}}{-1+3\sqrt{r}}, \quad r > 1, \\
\end{equation}
and
\begin{equation}
\label{eq:extISSO2}
r = 1, \quad 0 \leq \zmax \leq \frac{1}{1+\sqrt{2}}.\\
\end{equation}
These relations allow us to evaluate the $\Upsilon_z/\Upsilon_\psi$ on the ISSO in the extremal limit. We find that the maximum as attained when $z=\sqrt{2}-1$ --- i.e. $x =\sign{a} \sqrt{2(\sqrt{2}-1)}$ --- the turnover point for the two relations above. At this point the ratio becomes 2.

We thus find that the 2:1 $z\psi$-resonance is not accessible during any quasi-circular inspiral, and  the 3:2 $z\psi$-resonance is accessible only in a small part of the parameter space. The lowest order the $z\psi$-resonance that all quasi-circular inspirals must pass through is the 4:3. The location of this resonance relative to the ISSO is shown in Fig.~\ref{fig:zpsi43}. We see that this resonance happens around $1M$ before the ISSO, which should be well before the plunge phase for most inspirals.

We have here focused purely on the location of the $z\psi$-resonance in the orbital phase space. In order for something to happen at these resonances we also need dynamics that couple the corresponding frequency modes. In EMRI's the first order gravitational self-force is independent of the secondary's spin and therefore $\psi$ and hence cannot couple these modes. In~\cite{Ruangsri:2015cvg} it was argued that the Mathisson-Papapetrou-Dixon correction to the orbital dynamics also cannot provide such coupling at linear order in the spin of the smaller object. It is however likely that the first order self-torque (i.e. the linear in mass ratio correction to the dynamics of the secondary spin) will couple at these resonances, as will the second order self-force.

\section{Discussion}\label{sec:discussions}
This paper has provided an explicit closed-form analytic solution for the parallel transport equation along an arbitrary bound geodesic in Kerr spacetime. This result is useful as the zeroth order baseline for studying the dynamics of spinning secondaries and tides in extreme mass-ratio inspirals.

Alternatively, the solutions can be viewed as a providing a set of coordinates to describe EMRI dynamics, which facilitates a two timescale expansion. The phases $q_i$ provide the ``fast'' variable of the system varying on the orbital timescale, with all other variables varying on the inspiral timescale. Moreover, the phases $q_i$ are defined in such a way that at zeroth order there derivatives with respect to Mino time are independent of the phases themselves. This means that they provide the necessary starting position to systematically eliminate all variance on the orbital timescale from the perturbed equations of motion using near-identity transformations, allowing efficient integration of the evolution \cite{vandeMeent:2018rms}.

As an application of our analytical results, we studied the potential occurrence of $z\psi$-resonance in precessing quasi-circular EMRIs. We showed that only resonances with $1<\Omega_z/\Omega_\psi < 2$ can occur in such inspirals, with resonances with $1<\Omega_z/\Omega_\psi < \sqrt{2}$ occurring in every quasi-circular inspiral. A more detailed study of resonances involving both generic orbits and arbitrary integer combinations of $\Omega_r$, $\Omega_z$, and $\Omega_\psi$ will be left to future work.
   
Finally, as a bonus feature we provide an explicit closed for expression for the proper time along a bound geodesic in \ref{app:propertime}. This should of particular use for studying physics occurring predominantly in the local free falling frame following the geodesics, i.e. the study of tidal disruption of stars on highly eccentric orbits.

\begin{acknowledgements}
MvdM is grateful to Chris Kavanagh for help it checking the typesetting of the equations in this work. MvdM was supported by European Union's Horizon 2020 research and innovation programme under grant agreement No 705229. 
\end{acknowledgements} 

\section*{References} 
%\raggedright
\bibliography{../bib/journalshortnames,../bib/meent,../bib/commongsf,pt}

\appendix
\section{Constants of motion}\label{app:ELQ}
In order to allow readers to evaluate all formula's in this work without having to trawl through the literature we reproduce the analytic expressions for $\nE$, $\nL$, and $\nQ$ in terms of $\rmax$, $\rmin$, and $\zmax$ first given by \cite{Schmidt:2002qk}.

\begin{align}
\nE &= \sqrt{
	\frac{\kappa\rho+2\epsilon\sigma-2a\sqrt{\tfrac{\sigma}{a^2}\hh{\sigma\epsilon^2+\rho\epsilon\kappa-\eta\kappa^2}}}{\rho^2+4\eta\sigma}
},\\
\nL &= -\frac{g(\rmin)\nE-\sqrt{(g(\rmin)^2+h(\rmin)f(\rmin))\nE^2 -h(\rmin)d(\rmin)}}{h(\rmin)},\quad\text{and}\\
\nQ &= \zmax^2\hh{a^2(1-\nE^2)+\frac{\nL^2}{1-\zmax^2}}
\end{align}
with
\begin{align}
\kappa 		&:= d(\rmin)h(\rmax)- (\rmax\leftrightarrow\rmin),\\
\epsilon 	&:= d(\rmin)g(\rmax)- (\rmax\leftrightarrow\rmin),\\
\rho 		&:= f(\rmin)h(\rmax)- (\rmax\leftrightarrow\rmin),\\
\eta 		&:= f(\rmin)g(\rmax)- (\rmax\leftrightarrow\rmin),\\
\sigma 		&:= g(\rmin)h(\rmax)- (\rmax\leftrightarrow\rmin),
\end{align}
and
\begin{align}
d(r) &:=\Delta(r)\hh{r^2+a^2\zmax^2},\\
f(r) &:=r^4 + a^2\hh{r(r+2)+\zmax^2\Delta(r)},\\
g(r) &:=2ar,\quad\text{and}\\
h(r) &:=r(r-2)+\frac{\zmax^2\Delta(r)}{1-\zmax^2}.
\end{align}

%\section{Ellitpic Integrals}
%\begin{align}
%\d{\psi}{\lambda} = \sqrt{\nK}\hh{ \frac{(r^2+a^2)\nE-a\nL}{\nK+r^2} +a\frac{\nL-a (1-z^2)\nE}{\nK-a^2z^2}}.
%\end{align}
%
%\begin{align}
%\psi(\lambda) &= I_r(\lambda) + I_z(\lambda) + q_{\psi,0}\\
%I_r(\lambda) &= \sqrt{\nK}\int_0^{\lambda}\frac{(r(\hat\lambda)^2+a^2)\nE-a\nL}{\nK+r(\hat\lambda)^2}\id{\hat\lambda}
%&&= \sqrt{\nK}\oint_{\rmin}^{r(\lambda)}
% \frac{(r^2+a^2)\nE-a\nL}{(\nK+r^2)\d{r}{\lambda}}
% \id{r}
% \\
%I_z(\lambda) &= a\sqrt{\nK}\int_0^{\lambda}\frac{\nL-a (1-z(\hat\lambda)^2)\nE}{\nK-a^2z(\hat\lambda)^2}\id{\hat\lambda}
%&&= a\sqrt{\nK}\oint_0^{z(\lambda)}\frac{\nL-a (1-z^2)\nE}{(\nK-a^2z^2)\d{z}{\lambda}}\id{z}
%\end{align}

\section{Elliptic Functions}\label{app:elliptic}
We here give an overview of the definitions of the elliptic functions used throughout this paper. We generally follow the some notational conventions as {\sc Mathematica}.
{\centering
\begin{tabular}{ccc}
	\hline\hline
	Function	& 	Definition	& Description	\\
	\hline
	$\elF(\xi|k)$ 
		& $\displaystyle\int_0^\xi \frac{1}{\sqrt{1-k\sin^2\chi}}\id\chi$
		& \small Elliptic integral of the first kind
		\rule{0pt}{20pt}
		\\
	$\elK(k)$ 
		& $\elF(\tfrac{\pi}{2}|k)$
		& \small Complete elliptic integral of the first kind
		\\
	$\elE(\xi|k)$ 
		& $\displaystyle\int_0^\xi\sqrt{1-k\sin^2\chi}\id\chi$
		& \small Elliptic integral of the second kind
			\\
	$\elE(k)$ 
		& $\elE(\tfrac{\pi}{2}|k)$
		& \small Complete elliptic integral of the second kind\\
	$\elPi(h;\xi|k)$ 
		& $\displaystyle\int_0^\xi \frac{1}{(1-h\sin^2\chi)\sqrt{1-k\sin^2\chi}}\id\chi$
		& \small Elliptic integral of the third kind\\
	$\elPi(h|k)$ 
		& $\elPi(h; \tfrac{\pi}{2}|k)$
		& \small Complete elliptic integral of the third kind\\
	$\am(u|k)$
		&$ u = \elF\left(\am(u|k)\mid k\right)$
		&Jacobi amplitude\\		
	$\sn(u|k)$
		&$ \sin(\am(u|k))$
		&Jacobi elliptic sine\\		
	\hline\hline
\end{tabular}
}

\section{Proper time}\label{app:propertime}
The explicit closed form solutions for bound geodesics and parallel transport along them are given as functions of Mino time. However, for some applications it is useful to know the propertime along the orbit as well, e.g. when constructing  explicit local Fermi-Walker coordinates around the orbit.

The relation between Mino time and proper time is given by
\begin{equation}
\d{\tau}{\lambda} = \Sigma = r^2 + a^2 z^2.
\end{equation}
This equation can be solved through the same means as used to solve the equations for $t$, $\phi$, and $\psi$. We give the explicit solution here,
\begin{align}
\tau(q_\tau,q_r,q_z) &= q_\tau + \tau_r(q_r)+\tau_z(q_z),\\
\tau_r(q_r) &:=	\tilde{\tau}_r\Bh{\am\bh{\elK(k_r)\frac{q_r}{\pi}\big|k_r}} -\frac{\tilde{\tau}_r(\pi)}{2\pi}q_r,
\\
\tau_z(q_z) &:=	\tilde{\tau}_z\Bh{\am\bh{\elK(k_z)\frac{2q_z}{\pi}\big|k_z}}, -\frac{\tilde{\tau}_z(\pi)}{\pi}q_z
\end{align}
with
\begin{align}
\tilde{\tau}_r(\xi_r) &:= \frac{1}{\sqrt{(1-\nE^2)(\rmax-r_3)(\rmin-r_4)}}\Bh{
	\bh{(\rmax+\rmin+r_3)r_3-\rmax\rmin}\elF(\xi_r|k_r)
	\eqbreak
	+(\rmax-r_3)(\rmin-r_4)\elE(\xi_r|k_r)
	+(\rmin-r_3)(\rmax+\rmin+r_3+r_4)\elPi(h_r;\xi_r|k_r)
	\eqbreak
	-\frac{h_r (\rmax-r_3)(\rmin-r_4)\sin\xi_r\cos\xi_r \sqrt{1-k_r\sin^2\xi_r}}{1-h_r\sin^2\xi_r}
},\quad\text{and}
\\
\tilde{\tau}_z(\xi_z) &:=\frac{\zp}{(1-\nE^2)}\Bh{\elF(\xi_z|k_z)-\elE(\xi_z|k_z)}.
\end{align}
The secular part $q_\tau$ evolves linearly with $\lambda$,
\begin{equation}
q_\tau = \Upsilon_\tau \lambda +q_{\tau,0},
\end{equation}
with ``frequency''
\begin{align}
\Upsilon_{\tau} &= \tilde{\Upsilon}_{\tau,r}+\tilde{\Upsilon}_{\tau,z},\\
\tilde{\Upsilon}_{\tau,r} &:= \frac{(\rmax+\rmin+r_3)r_3-\rmax\rmin}{2}
+(\rmax-r_3)(\rmin-r_4)\frac{\elE(k_r)}{2\elK(k_r)}
\eqbreak
+(\rmax+\rmin+r_3+r_4)(\rmin-r_3)\frac{\elPi(h_r|k_r)}{2\elK(k_r)},\quad\text{and}\\
\tilde{\Upsilon}_{\tau,z} &:=\frac{\zp^2}{1-\nE^2}\hh{1-\frac{\elE(k_z)}{\elK(k_z)}}.
\end{align}
This explicit analytic expression for the $\Upsilon_\tau$ also allows for the calculation of the proper time frequencies $\omega_i$ through
\begin{equation}
\omega_i = \frac{\Upsilon_i}{\Upsilon_\tau}.
\end{equation}
\end{document}